# Gyrokinetic ion and drift kinetic electron model for electromagnetic simulation in the toroidal geometry

J. Bao[1], D. Liu[2], Z. Lin[1, *]

[1] Department of Physics and Astronomy, University of California, Irvine, California 92697, USA
[2] College of Physical Science and Technology, Sichuan University, Chengdu 610064, China
[*] Author to whom correspondence should be addressed. E-mail: *zhihongl@uci.edu*

The kinetic effects of electrons are important to long wavelength magnetohydrodynamic(MHD) instabilities and short wavelength drift-Alfvenic instabilities responsible for turbulence transport in magnetized plasmas, since the non-adiabatic electron can interact with, modify and drive the low frequency instabilities. A novel conservative split weight scheme is proposed for the electromagnetic simulation with drift kinetic electrons in tokamak plasmas, which shows great computational advantages that there is no numerical constrain of electron skin depth on the perpendicular grid size without sacrificing any physics. Both kinetic Alfven wave and collision-less tearing mode are verified by using this model, which has already been implemented into the gyrokinetic toroidal code(GTC). This model will be used for the micro tearing mode and neoclassical tearing mode simulation based on the first principle in the future.

## I. Introduction

With the fast developments of the parallel computing technique and numerical algorithm, the gyrokinetic particle simulation has been a powerful tool to help understand the various mechanisms of instability and transport in tokamak plasmas since three decades ago[1-7]. In particular, recent gyrokinetic particle simulation finds that the kinetic effects can affect the magneto-hydro-dynamic(MHD) modes by modifying the growth rates as well as mode structures, which brings us a more comprehensive sight on the electromagnetic physics compared to the perturbative theory[8-10]. On the other hand, to understand the neoclassical tearing mode physics is very important for preventing disruptions in tokamak[11], which requires the kinetic electrons for the first principle gyrokinetic simulation. However, compared to the electrostatic simulation, electromagnetic simulation with kinetic electrons encounters great computational challenges, especially for the long wavelength modes in high $\beta$ ( $\beta = 8\pi nT/B^2$ ) plasmas[12, 13]. The origin of this problem is that the large adiabatic part from the kinetic particles on one side is difficult to cancel with the corresponding term on the other side exactly in the parallel Ampere's law ( $p_{||}$ formulation) or generalized Ohm's law ($v_{||}$ formulation)[14, 15], which is sometimes referred as a "cancellation problem"[16, 17]. Any small error of the cancellation will cause the large error of the parallel vector potential in the high $\beta$ ($\beta_e \gg m_e/m_i$) and long wavelength ($k_\perp^2 d_e^2 \ll 1$, $k_\perp$ is the perpendicular wave vector and $d_e$ is the electron skin length) regime, which is true for the most instabilities in modern tokamak plasmas. Although many methods have been developed to overcome or avoid the "cancellation problem"[15-24], electromagnetic particle simulation with kinetic electrons is still computational challenging.

A fluid kinetic hybrid electron model[21-24] has been developed and implemented into Gyrokinetic Toroidal Code (GTC) for studying the electromagnetic instabilities in tokamak plasmas with high efficiency and accuracy[8, 26], which avoids the "cancellation problem". However, the hybrid model is based on the adiabatic expansion by assuming $\omega < k_{||}v_{the}$ ($\omega$ and $k_{||}$ are the wave frequency and parallel wave vector, and $v_{the} = \sqrt{T_e/m_e}$ is the electron thermal velocity), and it fails in the cases when the parallel wave-particle interaction is important near the rational surface, such as the tearing modes with $\omega/k_{||}v_{the} > 1$ in the inner region. Thus, it is urgent to incorporate the physics with tearing parity for the first principle study of neoclassical tearing mode and the disruption physics.

In this work, we have formulated and verified a new gyrokinetic ion and drift kinetic electron model for the first principle electromagnetic simulations. In this new model, we do not use the expansion method to solve the electric field by assuming $\omega < k_{||}v_{the}$. We apply a conservative split weight scheme to the drift kinetic electron with $v_{||}$ formulation by fixing the electron continuity equation and solving the Ohm's law for the non-adiabatic part of the parallel vector potential directly, and all the kinetic electron effects come into the system through the electron pressure term. Similar to the conservative scheme[27] for the drift kinetic electron with $p_{||}$ formulation, our new model does not suffer any inaccuracy problem or numerical instability as observed in the early numerical experiments[12, 13]. Most importantly, by using the new model, the perpendicular grid size does not need resolve the electron skin depth for numerical stability if the physical scale is larger than the electron skin depth. The dispersion relations of kinetic Alfven wave (KAW) in uniform plasmas and collision-less tearing mode in cylindrical geometry have been verified. This new model can faithfully capture the tearing mode physics, and it can be utilized for the first principle simulation of micro tearing mode[25] and neoclassical tearing mode in the future by adding the collision operator, which has already been developed in GTC code[28, 29].

The paper is organized as follows: The physics model for the electromagnetic simulations with drift kinetic electrons is introduced in Sec. II. The numerical results of the verifications of KAW and collision-less tearing mode are shown in Sec. III. Section 4 is the conclusion.

## II. Physics model

### A. Electromagnetic gyrokinetic particle simulation model

The gyrokinetic model has been widely used to study the low frequency waves and instabilities in tokamak plasmas. The following ordering is adopted in this paper:

$$\frac{\omega}{\Omega_{ci}} \sim \frac{\delta f}{F_0} \sim \frac{e\delta\phi}{T} \sim \frac{\delta B}{B_0} \sim \frac{k_{||}}{k_\perp} \sim \frac{1}{k_\perp L} \sim O(\varepsilon) \ll 1,$$

where $\omega$ and $\Omega_{ci}$ are the physical mode frequency and the ion cyclotron frequency, $\delta f$ and $F_0$ are the perturbed and equilibrium distributions, $\delta\phi$ and $\delta B$ are the perturbed electrostatic potential and perturbed magnetic field, $\rho_i = v_{thi}/\Omega_{ci}$ is the ion gyro radius, $v_{thi} = \sqrt{T_{i0}/m_i}$, $L$ is the plasma equilibrium scale length $L \sim (L_n = \nabla \ln n_{e0}, L_T = \nabla \ln T_{e0}, L_B = \nabla \ln B_0)$.

The gyrokinetic Vlasov equation describes the gyro-center dynamics by using gyrocenter position $\mathbf{R}$, magnetic moment $\mu$ and parallel velocity $v_{||}$ as independent variables in the five dimensional phase space, which is[30]:

$$\left(\frac{\partial}{\partial t} + \dot{\mathbf{R}} \cdot \nabla + \dot{v}_{||} \frac{\partial}{\partial v_{||}}\right) f_\alpha(\mathbf{R}, v_{||}, \mu, t) = 0, \tag{1}$$

$$\dot{\mathbf{R}} = v_{||} \mathbf{b_0} + v_{||} \frac{\langle \boldsymbol{\delta B} \rangle}{B_{||}^*} + \langle \mathbf{v_E} \rangle + \mathbf{v_c} + \mathbf{v_g}, \tag{2}$$

$$\dot{v}_{||} = -\frac{1}{m_\alpha} \frac{\mathbf{B}^*}{B_{||}^*} \cdot \left(Z_\alpha \nabla \langle \delta\phi \rangle + \mu \nabla B_0\right) - \frac{Z_\alpha}{cm_\alpha} \frac{\partial \langle \delta A_{||} \rangle}{\partial t}, \tag{3}$$

where $Z_\alpha$ and $m_\alpha$ represent the charge and mass of $\alpha = i, e$ particle species, respectively. $\delta\phi$ and $\delta A_{||}$ are the electrostatic potential and parallel vector potential, respectively. $\mathbf{B_0^*} = \mathbf{B_0} + (B_0 v_{||}/\Omega_{c\alpha}) \nabla \times \mathbf{b_0}$, $B_{||}^* = \mathbf{b_0} \cdot \mathbf{B_0^*}$, $\langle \boldsymbol{\delta B} \rangle = \nabla \times (\langle \delta A_{||} \rangle \mathbf{b_0})$ and $\mathbf{B}^* = \mathbf{B_0^*} + \langle \boldsymbol{\delta B} \rangle$. $\Omega_{c\alpha} = \frac{Z_\alpha B_0}{cm_\alpha}$ is the cyclotron frequency. $\langle \cdots \rangle = \frac{1}{2\pi} \int \mathbf{dx} d\xi \delta(\mathbf{R} + \boldsymbol{\rho_\alpha} - \mathbf{x})$ represents the gyro-average, $\xi$ is gyrophase angle, $\mathbf{x}$ is the particle position and $\boldsymbol{\rho_\alpha} = \frac{\mathbf{b_0} \times \mathbf{v_\perp}}{\Omega_{c\alpha}}$ is the gyro radius. To drift kinetic electron, the gyro-average can be dropped in Eqs. (1)-(3), and $\mathbf{R} = \mathbf{x}$ in drift kinetic limit. $\mathbf{v}_E$, $\mathbf{v}_c$ and $\mathbf{v}_g$ are perturbed $E \times B$ drift due to the electrostatic potential, magnetic curvature drift and magnetic gradient drift, respectively, which are given as:

$$\langle \mathbf{v}_E \rangle = \frac{c}{B_{\parallel}^*} \mathbf{b_0} \times \nabla \langle \delta\phi \rangle,$$

$$\mathbf{v}_c = \frac{c m_\alpha v_{\parallel}^2}{Z_\alpha B_{\parallel}^*} \mathbf{b_0} \times (\mathbf{b_0} \cdot \nabla \mathbf{b_0}),$$

and

$$\mathbf{v}_g = \frac{c \mu}{Z_\alpha B_{\parallel}^*} \mathbf{b_0} \times \nabla B_0.$$

The appearance of $B_{\parallel}^*$ ensures that the gyrocenter equations of motion preserve the Hamiltonian structure and satisfy the Liouville's theorem: $\frac{\partial B_{\parallel}^*}{\partial t} + \nabla \cdot \left( B_{\parallel}^* \dot{\mathbf{R}} \right) + \frac{\partial}{\partial v_{\parallel}} \left( \dot{v}_{\parallel} B_{\parallel}^* \right) = 0$ [30-32].

The electrostatic potential $\delta\phi$ is solved by the gyrokinetic Poisson's equation:

$$\frac{Z_i^2 n_{i0}}{T_{i0}} \left( \delta\phi - \delta\tilde{\phi} \right) = Z_i \bar{n}_i - e n_e, \tag{4}$$

where $\delta\tilde{\phi}(\mathbf{x},t) = \frac{1}{n_i} \int \mathbf{dv} f_i(\mathbf{R}, v_{\parallel}, \mu, t) \langle \delta\phi \rangle (\mathbf{R},t)$ is the second gyrophase-averaged potential, $\bar{n}_i(\mathbf{x},t) = \int \mathbf{dv} f_i(\mathbf{R}, v_{\parallel}, \mu, t)$ and $n_e(\mathbf{x},t) = \int \mathbf{dv} f_e(\mathbf{x}, v_{\parallel}, \mu, t)$ are the gyrophase-averaged ion and drift kinetic electron densities, $\int \mathbf{dv} = \frac{2\pi}{m_i} \int B_{\parallel}^* dv_{\parallel} d\mu \frac{1}{2\pi} \int \delta(\mathbf{R} + \boldsymbol{\rho}_\alpha - \mathbf{x}) d\mathbf{R} d\xi$.

The parallel vector potential $\delta A_{\parallel}$ is solved by the parallel Ampere's law:

$$\nabla_{\perp}^2 \delta A_{\parallel} = \frac{4\pi}{c} \left( \bar{J}_{\parallel i} + J_{\parallel e} \right), \tag{5}$$

where $\bar{J}_{\parallel i}(\mathbf{x},t) = Z_i \int v_{\parallel} \mathbf{dv} f_i(\mathbf{X}, v_{\parallel}, \mu, t)$ and $J_{\parallel e}(\mathbf{x},t) = q_e \int v_{\parallel} \mathbf{dv} f_e(\mathbf{x}, v_{\parallel}, \mu, t)$.

In principle, Eqs. (1)-(5) form a closed system of $v_{\parallel}$ formulation for electromagnetic simulations, which is also referred as "symplectic representation". The perpendicular Ampere's law can be added to incorporate the compressional magnetic perturbations in above model[Ge17]. However, it is difficult to calculate the time derivative of the parallel vector potential $\partial \delta A_{\parallel} / \partial t$ for $v_{\parallel}$ formulation by using a time difference method.

Theoretically, Eqs. (1)-(5) can be used for the full f simulation of gyrokinetic plasmas. However, due to the small value of the realistic electron-ion mass ratio and discrete particle noise, it is difficult to apply Eqs. (1)-(5) directly for the electromagnetic simulations in the tokamak geometry. Thus, the perturbative $\delta f$ simulation scheme[33, 34] and the split weight scheme[35, 36] are developed to decreasing the particle noises. In particular, the $\delta f$ scheme is good enough for simulating ion species with low particle noises in both electrostatic and electromagnetic cases[3, 8, 26], and the split weight scheme can help to improve the numerical performance dramatically in the electrostatic simulation with kinetic electrons[35]. However, when the split weight scheme is applied in the electromagnetic simulation with kinetic electrons, the so-called "cancellation problem" still appears in the field equations[36], and it has been claimed in some early studies that the perpendicular grid size should resolve the electron skin depth for an accurate cancellation if no other special numerical techniques are applied[13, 37].

In a previous work[27], it is shown that the electron perturbed density and current measured from kinetic markers do not satisfy the electron continuity equation in the conventional $\delta f$ scheme. Consequently, the electrostatic potential calculated from the density and the parallel vector potential calculated from the current are not consistent with each other, which results in an unphysically large parallel electric field. Instead of overcoming the "cancellation problem", the difficulty of electromagnetic simulation with kinetic electrons ($p_\parallel$ formulation) can be well solved by applying the electron continuity equation to time advance the electron density perturbation using the perturbed current calculated from the kinetic particles, which is referred as a "conservative scheme"[27]. For the $v_\parallel$ formulation, we need to fix both electron continuity and momentum equations, and the electron kinetic effects come into the system through the electron pressure. Based on this discovery, we extend the original fluid-kinetic hybrid electron model in gyrokinetic toroidal code (GTC) to the drift kinetic electron model with a conservative scheme.

## B. Perturbative δf simulation scheme

In order to minimize the discrete particle noise in the simulation, the perturbative $\delta f$ method is developed for both ion and electron species by splitting the total distribution function into the equilibrium and perturbed parts $f_\alpha = f_{\alpha 0} + \delta f_\alpha$. In the $\delta f$ method, only the perturbed distribution function $\delta f_\alpha$ is evolved in the simulation, which decreases the numerical noise of ions by a factor of $(\delta f_\alpha / f_\alpha)^2$. In the lowest order, the equation for the equilibrium distribution

$f_{\alpha 0}$ can be written as:

$$L_0 f_{\alpha 0} = 0,  \tag{7}$$

where $L_0 = \frac{\partial}{\partial t} + \left(v_{\parallel} \mathbf{b_0} + \mathbf{v_c} + \mathbf{v_g}\right) \cdot \nabla - \frac{\mu}{m_\alpha B_{\parallel}^*} \mathbf{B_0^*} \cdot \nabla B_0 \frac{\partial}{\partial v_{\parallel}}$ is the equilibrium propagator. $f_{\alpha 0}$ can be approximated as a shifted Maxwellian

$$f_{\alpha 0} = n_{\alpha 0} \left(\frac{m_\alpha}{2\pi T_{\alpha 0}}\right)^{3/2} \exp\left[-\frac{m_\alpha \left(v_{\parallel} - u_{\parallel \alpha 0}\right)^2 + 2\mu B}{2T_{\alpha 0}}\right],$$

where $u_{\parallel \alpha 0}$ is the parallel equilibrium flow velocity of each species, and $Z_i n_{i0} u_{\parallel i0} - e n_{e0} u_{\parallel e0} = \frac{c}{4\pi} \mathbf{b_0} \cdot \nabla \times \mathbf{B_0}$. A rigorous neo-classical solution which satisfies Eq. (7) can be constructed for simulation of the coupling between micro-turbulence and neo-classical transport.

Using Eq. (1) to subtract Eq. (7), the equation of $\delta f_\alpha$ can be derived as:

$$L \delta f_\alpha = -\left(\delta L_1 + \delta L_2\right) f_{\alpha 0}, \tag{8}$$

where $L = \frac{\partial}{\partial t} + \dot{\mathbf{X}} \cdot \nabla + \dot{v}_{\parallel} \frac{\partial}{\partial v_{\parallel}}$ is the total propagator, $\delta L_1$ and $\delta L_2$ are the linear and nonlinear perturbed propagators as:

$$\delta L_1 = \left(v_{\parallel} \frac{\langle \delta \mathbf{B} \rangle}{B_{\parallel}^*} + \langle \mathbf{v_E} \rangle\right) \cdot \nabla - \left[\frac{\mu}{m_\alpha B_{\parallel}^*} \langle \delta \mathbf{B} \rangle \cdot \nabla B_0 + \frac{Z_\alpha}{m_\alpha}\left(\frac{\mathbf{B_0^*}}{B_{\parallel}^*} \cdot \nabla \langle \delta \phi \rangle + \frac{1}{c} \frac{\partial \langle \delta A_{\parallel} \rangle}{\partial t}\right)\right] \frac{\partial}{\partial v_{\parallel}},$$

$$\delta L_2 = -\frac{Z_\alpha}{m_\alpha B_{\parallel}^*} \langle \delta \mathbf{B} \rangle \cdot \nabla \langle \delta \phi \rangle \frac{\partial}{\partial v_{\parallel}}.$$

In practice, we can rewrite Eq. (8) by defining the particle weight as $w_\alpha = \delta f_\alpha / f_\alpha$:

$$\frac{dw_\alpha}{dt} = (1 - w_\alpha) \left[\begin{array}{l} -\left(v_{\parallel} \frac{\langle \delta \mathbf{B} \rangle}{B_{\parallel}^*} + \langle \mathbf{v_E} \rangle\right) \cdot \frac{\nabla f_{\alpha 0}}{f_{\alpha 0}} \\ + \left(\mu \frac{\langle \delta \mathbf{B} \rangle}{B_{\parallel}^*} \cdot \nabla B_0 + Z_\alpha \frac{\mathbf{B^*}}{B_{\parallel}^*} \cdot \nabla \langle \delta \phi \rangle + \frac{Z_\alpha}{c} \frac{\partial \langle \delta A_{\parallel} \rangle}{\partial t}\right) \frac{1}{m_\alpha f_{\alpha 0}} \frac{\partial f_{\alpha 0}}{\partial v_{\parallel}} \end{array}\right]. \tag{9}$$

Eqs. (1)-(3) and (9) are applied to the $\delta f$ simulations.

## III. Conservative scheme for drift kinetic electron

In this subsection, we extend the fluid-kinetic hybrid electron model to the exact drift kinetic electron model by utilizing the conservative scheme[27] and the split weight scheme[35, 36]. In the following derivation of the electron equations in our model, we keep the terms up to the second order $O(\varepsilon^2)$ in the electron equations.

Similar to Eqs. (7) and (8), we can write the equations for the equilibrium and perturbed electron distributions as $f_e = f_{e0} + \delta f_e$:

$$L_0 f_{e0} = 0, \tag{10}$$

$$L\delta f_e = -(\delta L_1 + \delta L_2) f_{e0}, \tag{11}$$

where $f_{e0}$ could be a shifted Maxwellian. $L = L_0 + \delta L_1 + \delta L_2$, and $L_0$, $\delta L_1$ and $\delta L_2$ are the equilibrium, linear perturbed and nonlinear perturbed propagators for electron species, which are defined in Sec. II-B.

$\delta \tilde{f}_e$ is the $k_\parallel \neq 0$ component of the perturbed distribution $\delta f_e$, and it satisfies the following equation:

$$v_\parallel \mathbf{b_0} \cdot \nabla \delta \tilde{f}_e = -v_\parallel \frac{\boldsymbol{\delta B}^A}{B_0} \cdot \nabla f_{e0}\bigg|_\mu - \left[\frac{\mu}{B_0} \boldsymbol{\delta B}^A \cdot \nabla B_0 + q_e \mathbf{b_0} \cdot \nabla (\delta\phi + \delta\phi_{ind})\right] \frac{v_\parallel}{T_{e0}} f_{e0}. \tag{12}$$

where $\delta\phi_{ind}$ is only valid for the $k_\parallel \neq 0$ component, and is defined as:

$$\frac{\partial \delta A_\parallel^A}{\partial t} = c\mathbf{b_0} \cdot \nabla \delta\phi_{ind}. \tag{13}$$

The parallel vector potential $\delta A_\parallel$ consists of adiabatic and non-adiabatic parts as $\delta A_\parallel = \delta A_\parallel^A + \delta A_\parallel^{NA}$. $\boldsymbol{\delta B}^A$ is the adiabatic part of the magnetic perturbation, which is defined as

$$\boldsymbol{\delta B}^A = \nabla \times (\delta A_\parallel^A \mathbf{b_0}), \tag{14}$$

on the other hand, the adiabatic magnetic perturbation can be defined by using the Clebsch representation as:

$$\boldsymbol{\delta B}^A = \nabla \psi_0 \times \nabla \delta\alpha^A + \nabla \delta\psi^A \times \nabla \alpha_0 \tag{15}$$

where $\psi = \psi_0 + \delta\psi$ is the poloidal flux label, $\theta$ and $\zeta$ are poloidal and toroidal angles in magnetic coordinates, and $\alpha = q(\psi)\theta - \zeta$ is the magnetic field line label.

Using Eq. (15), we can solve Eq. (12) and derive the solution of $\delta \tilde{f}_e$:

$$\delta \tilde{f}_e = \frac{e(\delta\phi + \delta\phi_{ind})}{T_{e0}} f_{e0} + \left.\frac{\partial f_{e0}}{\partial \psi_0}\right|_{v_\perp} \delta\psi^A + \left.\frac{\partial f_{e0}}{\partial \alpha_0}\right|_{v_\perp} \delta\alpha^A - \frac{e\langle\langle\delta\phi\rangle\rangle}{T_{e0}} f_{e0}. \qquad (16)$$

Integrating Eq. (16) to the zeroth order, we have:

$$\frac{e\delta\phi_{ind}}{T_{e0}} = \frac{\delta n_e}{n_{e0}} - \frac{e\delta\phi}{T_{e0}} - \frac{\partial n_{e0}}{\partial \psi} \frac{\delta\psi^A}{n_{e0}} - \frac{\partial n_{e0}}{\partial \alpha} \frac{\delta\alpha^A}{n_{e0}} - \left( \frac{\langle\langle\delta n_e\rangle\rangle}{n_{e0}} - \frac{e\langle\langle\delta\phi\rangle\rangle}{T_{e0}} \right), \qquad (17)$$

where $\delta n_e = \int \delta f_e \mathbf{dv}$, $\delta n_e(k_{\parallel} \neq 0) = \int \delta \tilde{f}_e \mathbf{dv} = \delta n_e - \langle\langle \delta n_e \rangle\rangle$, $\langle\langle \cdots \rangle\rangle = \dfrac{\int (\cdots) J d\theta d\zeta}{\int J d\theta d\zeta}$

represents the flux surface averaging, and $J$ is Jacobian.

From Eqs. (14) and (15), we obtain the equations for $\delta\psi^A$ and $\delta\alpha^A$:

$$\frac{\partial \delta\psi^A}{\partial t} = -c \frac{\partial \delta\phi_{ind}}{\partial \alpha_0}, \qquad (18)$$

$$\frac{\partial \delta\alpha^A}{\partial t} = c \frac{\partial \delta\phi_{ind}}{\partial \psi_0}, \qquad (19)$$

It should be noted that $\delta A_{\parallel}^A$, $\delta\phi_{ind}$, $\delta\psi^A$, $\delta\alpha^A$ are only valid for the $k_{\parallel} \neq 0$ components, namely, $\delta A_{\parallel}^A = \delta A_{\parallel}^A(k_{\parallel} \neq 0)$, $\delta\phi_{ind} = \delta\phi_{ind}(k_{\parallel} \neq 0)$, $\delta\psi^A = \delta\psi^A(k_{\parallel} \neq 0)$, $\delta\alpha^A = \delta\alpha^A(k_{\parallel} \neq 0)$.

Next we take the moment of Eq. (11) to obtain the equation for $\delta n_e$ as:

$$\frac{\partial \delta n_e}{\partial t} + \nabla \cdot \left[ n_{e0} \left( \delta u_{\parallel e} \mathbf{b_0} + \mathbf{V_E} + u_{\parallel e0} \frac{\delta \mathbf{B}}{B_0} \right) + \frac{1}{T_{e0}} \left( \delta P_{\perp e} \mathbf{V_g} + \delta P_{\parallel e} \mathbf{V_c} \right) + \frac{n_{e0} \delta u_{\parallel e}}{B_0} \delta\mathbf{B} + \delta n_e \mathbf{V_E} \right] = 0, \qquad (20)$$

where $\delta u_{\parallel e}$ and $u_{\parallel e0}$ are the perturbed and equilibrium parallel velocities of electron gyrocenter, and $\delta P_{\parallel e}$ and $\delta P_{\perp e}$ are the perturbed parallel and perpendicular pressures of electron gyrocenter. $\mathbf{V_E} = c\mathbf{b_0} \times \nabla \delta\phi / B_0$ is $E \times B$ drift, $\mathbf{V_c} = \int \mathbf{v_c} f_{e0} \mathbf{dv} = \dfrac{cT_{e0}}{q_e B_0} \mathbf{b_0} \times (\mathbf{b_0} \cdot \nabla \mathbf{b_0})$, and $\mathbf{V_g} = \int \mathbf{v_g} f_{e0} \mathbf{dv} = \dfrac{cT_{e0}}{q_e B_0^2} \mathbf{b_0} \times \nabla B_0$.

On the other hand, the perturbed electron distribution consists of adiabatic and non-adiabatic part as $\delta f_e = \delta f_a + \delta h$. The adiabatic electron response $\delta f_a$ is defined to satisfy the following equation:

$$v_{\|}\mathbf{b_0} \cdot \nabla \delta f_a = -v_{\|} \frac{\delta \mathbf{B}^A}{B_0} \cdot \nabla f_{e0}\bigg|_{\mu} - \left[\frac{\mu}{B_0}\delta \mathbf{B}^A \cdot \nabla B_0 + q_e \mathbf{b_0} \cdot \nabla \left(\delta \phi + \delta \phi_{ind}\right)\right]\frac{v_{\|}}{T_{e0}} f_{e0}. \tag{21}$$

By using the relation of Eq. (15), we can solve Eq. (21) for $\delta f_a$:

$$\delta f_a = \frac{e\left(\delta \phi + \delta \phi_{ind}\right)}{T_{e0}} f_{e0} + \frac{\partial f_{e0}}{\partial \psi_0}\bigg|_{v_\perp} \delta \psi^A + \frac{\partial f_{e0}}{\partial \alpha_0}\bigg|_{v_\perp} \delta \alpha^A - \frac{e\langle\langle\delta\phi\rangle\rangle}{T_{e0}} f_{e0}. \tag{22}$$

Substituting $\delta f_a$ into Eq. (11), the equation of non-adiabatic electron response can be obtained as:

$$L\delta h_e = \underbrace{-\delta L_1 f_{e0} - L_0 \delta f_a}_{\{I\}} \underbrace{-\delta L_2 f_{e0} - \left(\delta L_1 + \delta L_2\right)\delta f_a}_{\{II\}}, \tag{23}$$

where term {I} is linear and term {II} is nonlinear.

We define a weight function for the non-adiabatic electron response as: $w_e = \delta h / f_e$ and one can write the electron weight equation from Eq. (23):

$$\frac{dw_e}{dt} = \frac{1-w_e}{1+\delta f_a/f_{e0}} \frac{1}{f_{e0}}\left[-\delta L_1 f_{e0} - L_0 \delta f_a - \delta L_2 f_{e0} - \left(\delta L_1 + \delta L_2\right)\delta f_a\right] \tag{24}$$

where we keep the terms up to the second order $O\left(\varepsilon^2\right)$ in the square bracket on the RHS. By using the relations of Eqs. (17) and (22), the time derivative of the adiabatic response $\partial \delta f_a/\partial t$ in Eq. (24) can be calculated as:

$$\frac{1}{f_{e0}}\frac{\partial \delta f_a}{\partial t} = \frac{1}{n_{e0}}\frac{\partial \delta n_e}{\partial t} - \frac{1}{n_{e0}}\frac{\partial \langle\langle\delta n_e\rangle\rangle}{\partial t} + \frac{1}{f_{e0}}\frac{\partial f_{e0}}{\partial T_{e0}}\frac{\partial T_{e0}}{\partial \psi_0}\bigg|_{v_\perp}\frac{\partial \delta \psi^A}{\partial t} + \frac{1}{f_{e0}}\frac{\partial f_{e0}}{\partial T_{e0}}\frac{\partial T_{e0}}{\partial \alpha_0}\bigg|_{v_\perp}\frac{\partial \delta \alpha^A}{\partial t}. \tag{25}$$

The equation for the non-adiabatic vector potential $\delta A_{\|}^{NA}$ is:

$$\left(\nabla_\perp^2 - \frac{\omega_{pe}^2}{c^2}\right)\frac{\partial \delta A_{\|}^{NA}}{\partial t} = \frac{\omega_{pe}^2}{c^2}\chi_{\||e} - c\nabla_\perp^2\left(\mathbf{b_0}\cdot\nabla\delta\phi_{ind}\right), \tag{26}$$

where

$$\chi_{\|e} = \underbrace{-\frac{c}{en_{e0}}\mathbf{b_0}\cdot\nabla\delta P_{\|e}^{NA}}_{\{I\}} \underbrace{-\frac{c}{en_{e0}B_0}\delta\mathbf{B}^{NA}\cdot\nabla P_{\|e0}}_{\{II\}} \underbrace{-\frac{c}{B_0}\delta\mathbf{B}\cdot\nabla\delta\phi_{ind}}_{\{III\}} \underbrace{-\frac{c}{en_{e0}B_0}\delta\mathbf{B}\cdot\nabla\delta P_{\|e}^{NA}}_{\{IV\}}$$

$$\underbrace{-\frac{cm_e}{en_{e0}}\nabla\cdot\left[n_{e0}\delta u_{\|e}\left(3\mathbf{V_c}+\mathbf{V_g}\right)+n_{e0}u_{\|e0}\mathbf{V_E}\right]}_{V} \underbrace{-\frac{cm_e}{en_{e0}}\nabla\cdot\left(n_{e0}\delta u_{\|e}\mathbf{V_E}\right)}_{VI},$$

$$\underbrace{+\frac{c}{en_{e0}}\frac{P_{\|e0}-P_{\perp e0}}{B_0^2}\delta\mathbf{B}\cdot\nabla B_0 + \frac{c}{en_{e0}}\frac{\delta P_{\|e}^{NA}-\delta P_{\perp e}^{NA}}{B_0^2}\mathbf{B_0}\cdot\nabla B_0}_{VII}$$

and we drop the ion contribution $\partial\delta\overline{u}_{i\|}/\partial t$ due to the small electron-ion mass ratio in Eq. (26). Terms {I}, {II}, {V} and {VII} are linear, and terms {III}, {IV} and {VI} are nonlinear. The derivation of Eq. (26) is shown in the Appendix A.

It requires the values of $\delta u_{\|e}$, $\delta P_{\|e}$ and $\delta P_{\perp e}$ in order to evolve Eqs. (20) and (21).

$\delta u_{\|e}$ can be calculated by inverting the Ampere's law Eq. (5) as:

$$\delta u_{\|e} = \frac{c}{4\pi en_{e0}}\nabla_\perp^2 \delta A_{\|} + \frac{Z_i}{e}\delta\overline{u}_{\|i}. \tag{27}$$

The electron parallel and perpendicular pressures consists of adiabatic and non-adiabatic parts as:

$$\delta P_{\|e} = \int_{DK} m_e v_\|^2 \left(\delta f_a + \delta h\right)\mathbf{dv} = \delta P_{\|e}^A + \delta P_{\|e}^{NA},$$

$$\delta P_{\perp e} = \int_{DK} \mu B_0 \left(\delta f_a + \delta h\right)\mathbf{dv} = \delta P_{\perp e}^A + \delta P_{\perp e}^{NA}.$$

Then we take the moment of Eq. (22) to get the adiabatic electron pressures as:

$$\delta P_{\|e}^A = en_{e0}\left(\delta\phi + \delta\phi_{ind}\right) + \frac{\partial(n_{e0}T_{e0})}{\partial\psi_0}\delta\psi^A + \frac{\partial(n_{e0}T_{e0})}{\partial\alpha_0}\delta\alpha^A - en_{e0}\langle\langle\delta\phi\rangle\rangle, \tag{28}$$

$$\delta P_{\perp e}^A = en_{e0}\left(\delta\phi + \delta\phi_{ind}\right) + \frac{\partial(n_{e0}T_{e0})}{\partial\psi_0}\delta\psi^A + \frac{\partial(n_{e0}T_{e0})}{\partial\alpha_0}\delta\alpha^A - en_{e0}\langle\langle\delta\phi\rangle\rangle. \tag{29}$$

The non-adiabatic electron pressures are calculated by using the non-adiabatic electron response $\delta h_e$:

$$\delta P_{\|e}^{NA} = \int m_e v_\|^2 \delta h \mathbf{dv}, \tag{30}$$

$$\delta P_{\perp e}^{NA} = \int \mu B_0 \delta h \mathbf{dv}. \tag{31}$$

Eqs. (1)-(4), (9), (13)-(15), (17)-(20), (24)-(31) form a closed system for electromagnetic simulation with kinetic electrons. It can be seen that all the electron kinetic effects come into our system from the perturbed kinetic pressures, which guarantees the conservation properties of electron perturbed density, parallel velocity and pressures through Eqs. (20) and (26). It will be shown later that this model does not suffer the well-known cancellation problem without resolving the electron skin depth.

## III. Numerical results

To verify this new model, we implement it into gyrokinetic toroidal code (GTC), and carry out the simulations of KAW in uniform plasmas and the collision-less tearing mode in the cylindrical geometry with magnetic shear.

### A. Kinetic Alfven wave in uniform plasmas

In uniform and linear plasmas, by assuming ion only provides the polarization density, we first use this new model to derive the dispersion relation. Eq. (23) is physically equivalent to Eq. (24), and in linear and uniform plasmas Eq. (23) can be written as:

$$\left(\frac{\partial}{\partial t}+v_{\parallel}\mathbf{b_0}\cdot\nabla\right)\delta h_e = -\frac{e}{m_e}\left(\mathbf{b_0}\cdot\nabla\delta\phi+\frac{1}{c}\frac{\partial\delta A_{\parallel}}{\partial t}\right)\frac{\partial f_{e0}}{\partial v_{\parallel}} - \left(\frac{\partial}{\partial t}+v_{\parallel}\mathbf{b_0}\cdot\nabla\right)\delta f_a. \tag{D1}$$

In the long wavelength limit, the gyrokinetic Poisson's equation (4) can reduce to

$$\frac{c^2}{4\pi e V_A^2}\nabla_{\perp}^2\delta\phi = \delta n_e. \tag{D2}$$

In uniform and linear plasmas, to $k_{\parallel}\neq 0$ modes, Eqs. (17), (20), (22), (26) and (27) reduce to

$$en_{e0}\left(\delta\phi+\delta\phi_{ind}\right) = \delta n_e T_{e0}, \tag{D3}$$

$$\frac{\partial\delta n_e}{\partial t}+n_{e0}\mathbf{b_0}\cdot\nabla\delta u_{\parallel e} = 0, \tag{D4}$$

$$\delta f_a = \frac{e\left(\delta\phi+\delta\phi_{ind}\right)}{T_{e0}}f_{e0}, \tag{D5}$$

$$\left(\nabla_{\perp}^2-\frac{\omega_{pe}^2}{c^2}\right)\frac{\partial\delta A_{\parallel}^{NA}}{\partial t} = -c\nabla_{\perp}^2\left(\mathbf{b_0}\cdot\nabla\phi_{ind}\right)-\frac{\omega_{pe}^2}{en_{e0}c}\mathbf{b_0}\cdot\nabla\delta P_{\parallel e}^{NA}. \tag{D6}$$

$$\delta u_{\parallel e} = \frac{c}{4\pi e n_{e0}}\nabla_{\perp}^2\delta A_{\parallel}, \tag{D7}$$

Applying the Fourier transform to Eqs. (D1)-(D7) and (13): $\partial_t\rightarrow -i\omega$, $\mathbf{b_0}\cdot\nabla\rightarrow ik_{\parallel}$ and $\nabla_{\perp}\rightarrow i\mathbf{k}_{\perp}$, the linear dispersion relation of KAW based on this model in the uniform plasmas is:

$$\left(\frac{\omega^2}{k_{\parallel}^2 V_A^2}-1\right)\left[1+\xi_e Z(\xi_e)\right] = k_{\perp}^2\rho_s^2, \tag{D8}$$

where $\xi_e = \omega/\sqrt{2}k_\parallel v_{the}$, $\rho_s = C_s/\Omega_{ci}$, $v_{the} = \sqrt{T_{e0}/m_e}$, $C_s = \sqrt{T_{e0}/m_i}$, and $Z(\xi_e)$ is the plasma dispersion function: $Z(\xi_e) = \frac{1}{\sqrt{\pi}}\int_{-\infty}^{+\infty}\frac{e^{-t^2}}{t-\xi_e}dt$. The effects of the compressional magnetic perturbation on the KAW are also studied by using GTC code recently, which can be easily added to this model[Ge2017].

In the simulation of KAWs, the electron temperature $T_{e0} = 5.0 keV$ and magnetic field $B_0 = 1.5T$ are uniform, the ratio between the parallel and perpendicular wave vectors is fixed as $k_\parallel/k_\perp = 0.01$, and the electron density changes as well as the value of $\beta_e = 8\pi n_{e0}T_{e0}/B_0^2$. Firstly we set $n_{e0} = 1.0\times 10^{13} cm^{-3}$ and the corresponding $\beta_e = 0.9\%$, and verify the dependence of the frequency on the wavelength $k\rho_s$ as well as the perpendicular grid size $\Delta x/d_e$. The simulation results agree well with the theory as shown by Fig. 1, and it can be seen that the perpendicular grid size does not need to resolve the electron skin length for the accuracy. Secondly, we fix the wavelength $k\rho_s = 0.48$, and change the electron density from $n_{e0} = 1.0\times 10^{13} cm^{-3}$ to $n_{e0} = 2.0\times 10^{14} cm^{-3}$, and verify the dispersion relations of KAW for different $\beta_e$ values. As shown by Fig. 2, both the frequency and damping rate agree with theory very well when $\beta_e \gg m_e/m_i$.

From the benchmark results, it can be seen that there is no constrain of the electron skin depth to the perpendicular grid size by using our model, and the simulation accuracy can be achieved in both long wavelength and high $\beta_e$ regimes, which shows great advantages of this model.

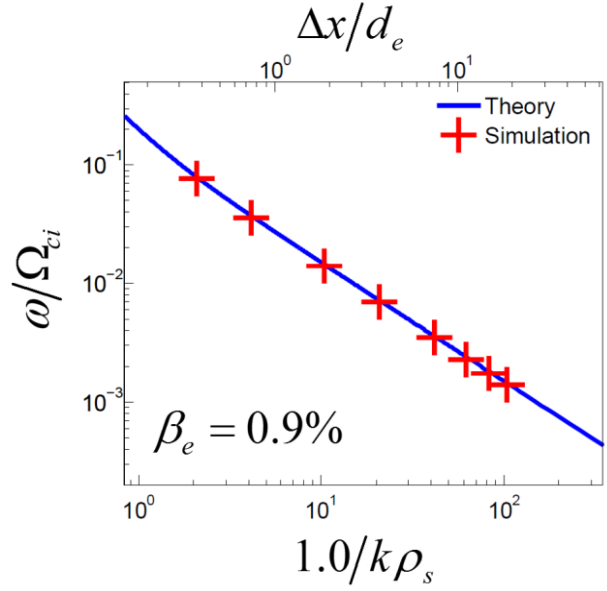

**FIG. 1.** The dependence of KAW frequency on the wavelength (bottom) and the perpendicular grid size (top).

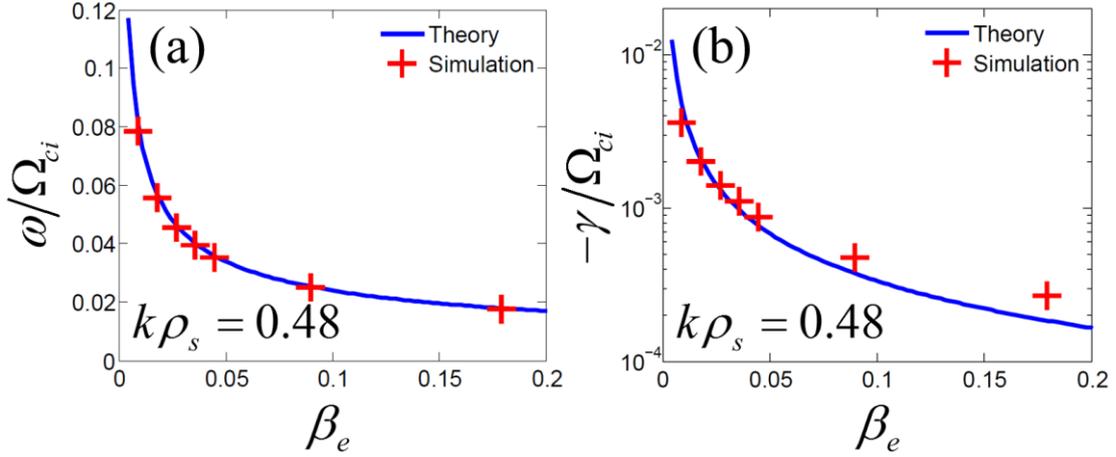

**FIG. 2.** The dependence of the KAW frequency and damping rate on $\beta_e$ (unit 100%). In high $\beta_e$ regime, the damping of KAW is too weak to measure, since the ion Landau damping is removed in the simulation, which dominates instead of electron Landau damping in high $\beta_e$ regime.

## B. Collision-less tearing mode in cylindrical geometry

In this part, we verify the collision-less tearing mode in both theory and simulation by using our model. Here, we only keep the tearing mode driven term of equilibrium current for simplicity, and neglect the equilibrium density and temperature gradient which contributes to the real frequency of tearing mode. Then we would get purely growing collision-less tearing mode using the following simplified model. From Eqs. (D1), (D3), (D7), (13) and (30), one can readily have the wave equation for tearing mode, which is:

$$\nabla_\perp^2 \omega \delta A_{||} = \frac{\omega_{pe}^2}{c}(\frac{1}{c}\omega \delta A_{||} - k_{||}\delta\phi)\xi_e^2 Z'(\xi_e), \qquad (D9)$$

where $Z'(\xi_e)$ is the derivative of the plasma function with respect to $\xi_e$.

For the macro tearing mode in the plasmas with uniform density and temperature profiles but non-uniform magnetic field, the continuity equation Eq. (20) reduces to

$$\frac{\partial \delta n_e}{\partial t} + n_{e0}\mathbf{b_0}\cdot\nabla\delta u_{||e} + n_{e0}\frac{\delta\mathbf{B}}{B_0}\cdot\nabla u_{||e0} = 0, \qquad (D10)$$

where $u_{||e0} = -\frac{c}{4\pi e n_{e0}}\mathbf{b_0}\cdot\nabla\times\mathbf{B_0}$ by neglecting the ion equilibrium flow.

Combining the Eqs. (D2), (D6), (D9) and (D10), one will have:

$$\nabla_\perp^4 \omega^2 \delta A_{||} = \nabla_\perp^2[\frac{1}{d_e^2}\omega^2 \xi_e^2 Z'(\xi_e)\delta A_{||}] - \frac{m_i}{m_e}k_{||}\xi_e^2 Z'(\xi_e)(k_{||}\nabla_\perp^2 - k_{||}'')\delta A_{||}, \qquad (D11)$$

where $k_{||}'' = d^2 k_{||}/dr^2$.

Following that of Drake and Lee[38], also Liu and Chen[39], one can have the inner region equation for the collision-less tearing mode:

$$\nabla_\perp^2 \delta A_{||} = \frac{1}{d_e^2}\xi_e^2 Z'(\xi_e)\delta A_{||}. \qquad (D12)$$

If one has the analytical solution $\delta A_{||o}$ for the outer region equation:

$$(k_{||}\nabla_\perp^2 - k_{||}'')\delta A_{||} = 0. \qquad (D13)$$

Using the constant $\delta A_{||}$ approximation, and matching the inner and outer region by using the boundary condition:

$$\frac{\delta\hat{A}_{||i}'}{\delta\hat{A}_{||i}} = \frac{\delta\hat{A}_{||o}'}{\delta\hat{A}_{||o}} = \Delta'. \qquad (D14)$$

One will have the dispersion relation for collision-less tearing mode as:

$$-i\omega = \gamma = \frac{d_e^2}{\sqrt{\pi}}|k_{||}'v_{te}|\Delta_o'. \qquad (D15)$$

Eq. (D15) is the same with the result from Drake and Lee[38].

When we set $\delta P_{||e}^{NA} = 0$ in Eq. (D7) and remain other equations to derive the dispersion relation, we can get the growth rate as:

$$-i\omega = \gamma = \frac{d_e^2}{\pi}|k_{\parallel}'v_{te}|\Delta_o'. \tag{D16}$$

Eq. (D16) is the same with the result which is derived from D. Liu and L. Chen's fluid model[39].

In the simulation, the equilibrium parameters are given as following: uniform equilibrium electron density $n_{e0} = 1.0 \times 10^{12} cm^{-3}$ and uniform electron temperature $T_{e0} = 5.0 keV$, on axis toroidal magnetic field $B_0 = 1.0T$, and major radius $R_0 = 1.0m$. It should be pointed out that the radial grid size need to resolve the electron skin depth $d_e$ near the rational surface for the collision-less tearing mode simulation due to the physics requirements, which is different from the KAW simulation. In the cylindrical geometry, we carry out the simulations of the collision-less tearing mode in both fluid and kinetic regimes by using our model with the q profile as shown by Fig. 3. Firstly, we drop the second term on the RHS of Eq. (D7), and our model reduces to the finite mass fluid electron model[39, 40]. The fluid simulation with the realistic electron-ion mass ratio $m_e/m_i = 1/1837$ gives us the growth rate $\gamma = 0.0014(C_s/R_0)$, and the mode structures of the parallel vector potential $\delta A_{\parallel}$ and electrostatic potential $\delta\phi$ on the poloidal plane are shown in Fig.4 (a) and (b). At the same time, we use an 1D eigenvalue code[39] to calculate the growth rate and mode structure in the fluid limit, which gives $\gamma = 0.0015(C_s/R_0)$, and GTC fluid simulation result agrees with eigenvalue result closely. Secondly, we apply the exact Eq. (D7) with the non-adiabatic pressure term in the kinetic electron simulation. By using the same equilibrium parameter, the kinetic simulation gives us the growth rate of the collision-less tearing mode $\gamma = 0.0031(C_s/R_0)$, and the mode structures of the parallel vector potential $\delta A_{\parallel}$ and electrostatic potential $\delta\phi$ on the poloidal plane are shown in Fig.4 (c) and (d). The theoretical growth rate of kinetic electron model is estimated from the fluid eigenvalue result by using Eqs. (D15) and (D16), which gives $\gamma = 0.0027(C_s/R_0)$, and GTC kinetic simulation result agree with the theory closely. Finally, we compare the radial mode structure of $\delta A_{\parallel}$ between the fluid eigenvalue, GTC fluid and GTC kinetic simulation results as shown in Fig. 5, and it can be seen that the GTC fluid simulation result agrees well with the eigenvalue result, and the kinetic effects of electron do not affect the radial mode structure.

From above simulation results, we can see that the linear growth rate of collision-less tearing mode in kinetic electron simulation is roughly $\sqrt{\pi}$ times that in fluid simulation, which has been predicted by the theory[38, 39]. The radial mode structures of $\delta A_{\parallel}$ from GTC simulations also agree with 1D eigenvalue calculation. Thus, our model can faithfully capture the physics of collision-less tearing mode in both fluid and kinetic regimes.

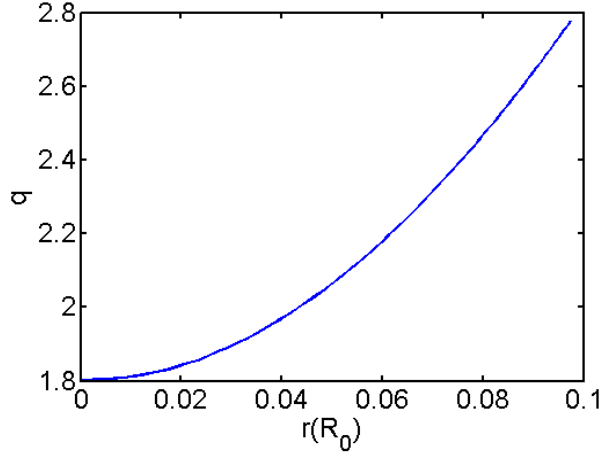

**FIG. 3**. The safety factor q profile for the collision-less tearing mode simulation.

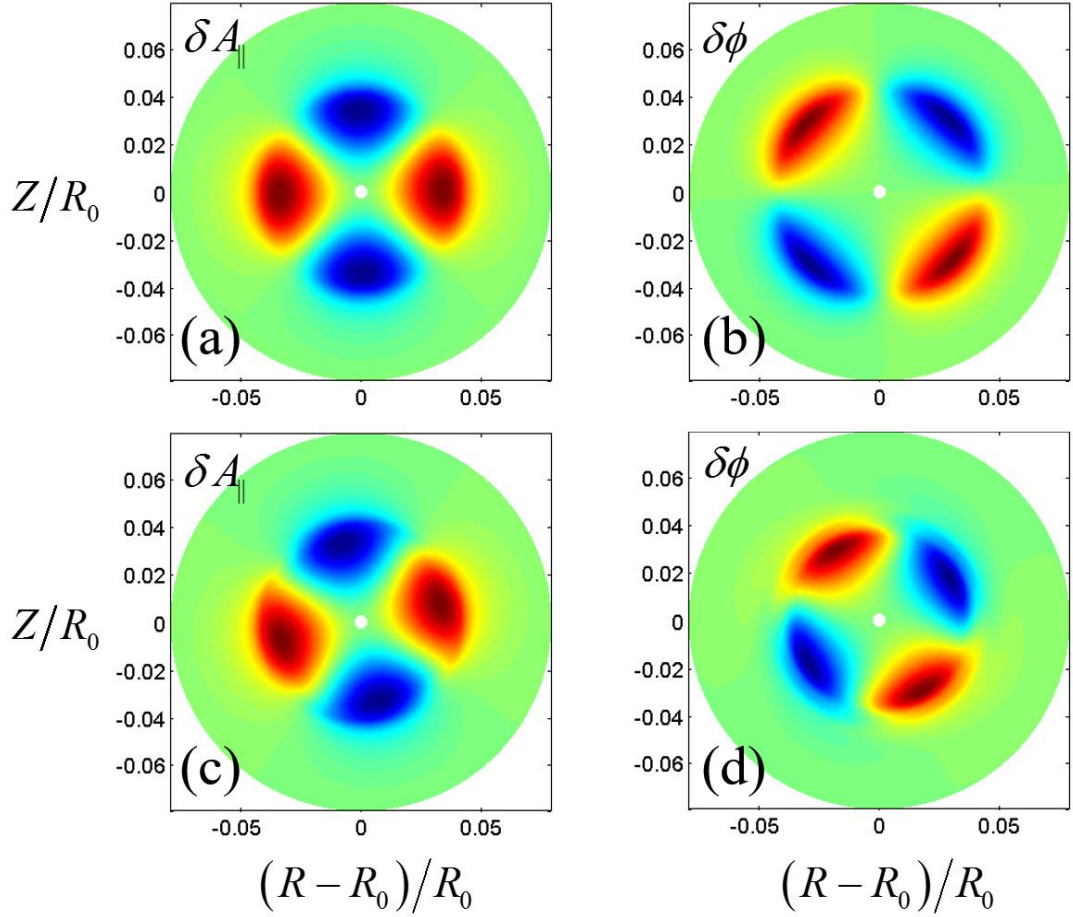

**FIG. 4.** Upper panel shows the mode structures of (a) $\delta A_\parallel$ and (b) $\delta\phi$ from the fluid electron simulation of the (2,1) collision-less tearing mode in the cylindrical geometry. Lower panel shows the mode structures of (c) $\delta A_\parallel$ and (d) $\delta\phi$ from the kinetic electron simulation of the (2,1)

collision-less tearing mode in the cylindrical geometry.

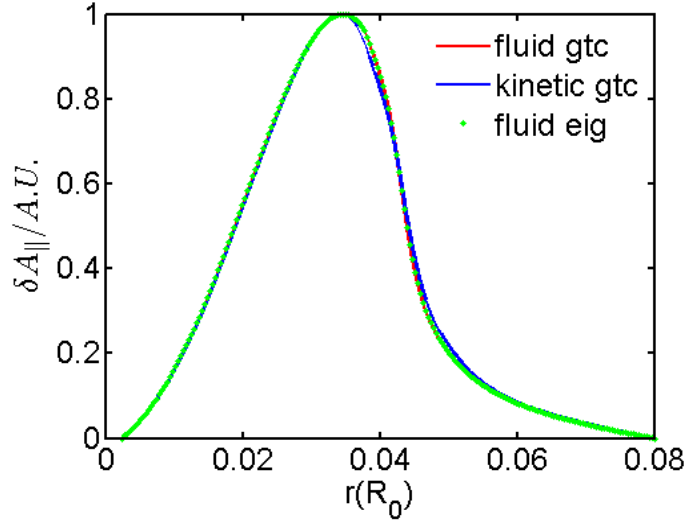

**FIG. 5.** Comparison of the radial mode structure of the (2, 1) collision-less tearing mode between GTC fluid simulation, GTC kinetic simulation and fluid eigenvalue calculation in the cylindrical geometry.

## IV. Conclusions

In this work, we present an efficient gyrokinetic ion and drift kinetic electron model for the electromagnetic simulations in the toroidal geometry. Gyrokinetic ion is simulated by using a conventional $\delta f$ scheme, and a novel conservative split weight scheme is applied to the drift kinetic electron, which guarantees the conservation properties of electron perturbed density, parallel velocity and pressures through the electron continuity equation and generalized Ohm's law. The dispersion relation of KAW is verified in the uniform plasmas, and it shows that the perpendicular grid size does not need to resolve the electron skin depth for the numerical accuracy and stability when the physical scale is larger than the electron skin depth. Finally, both the growth rate and mode structure of the collision-less tearing mode are verified with the theory and eigenvalue calculation, and this model can be utilized for the simulation of micro tearing mode and neoclassical tearing mode in the future.

## Acknowledgements

We would like to thank W. W. Lee, W. M. Tang, Y. Chen, I. Holod, L. Shi and Z. X. Lu for useful discussions. This work was supported by China National Magnetic Confinement Fusion Science Program (Grant No. 2013GB111000), US Department of Energy (DOE) SciDAC GSEP Program. This work used resources of the Oak Ridge Leadership Computing Facility at Oak Ridge National Laboratory (DOE Contract No. DE-AC05-00OR22725) and the National Energy Research Scientific Computing Center (DOE Contract No. DE-AC02-05CH11231).

# Appendix A: Derivation of the equation for non-adiabatic parallel vector potential

Integrating the electron momentum equation by using Eq. (11), one can write the momentum equation for electron species as:

$$n_{e0}\frac{\partial \delta u_{\parallel e}}{\partial t} + \nabla \cdot \left[ n_{e0}\delta u_{\parallel e}\left(\mathbf{V_E} + 3\mathbf{V_c} + \mathbf{V_g}\right) + n_{e0}u_{\parallel e0}\mathbf{V_E}\right]$$
$$+ \frac{q_e}{m_e}n_{e0}\left[\left(\mathbf{b_0} + \frac{\delta \mathbf{B}}{B_0}\right)\cdot \nabla \delta\phi + \frac{1}{c}\frac{\partial \delta A_{\parallel}}{\partial t}\right] + \frac{1}{m_e}\nabla \cdot \left[\delta P_{\parallel e}\left(\mathbf{b_0} + \frac{\delta \mathbf{B}}{B_0}\right) + P_{\parallel e0}\frac{\delta \mathbf{B}}{B_0}\right]. \quad (A1)$$
$$+ \frac{1}{m_e}\left(\frac{P_{\perp e0}}{B_0^2}\delta \mathbf{B}\cdot \nabla B_0 + \frac{\delta P_{\perp e}}{B_0}\mathbf{b_0}\cdot \nabla B_0\right) = 0$$

In the derivation of Eq. (A1), we only keep the terms up to the second order $O(\varepsilon^2)$ and make the truncations on the moments: $\int m_e v_{\parallel}^3 \delta f_e \mathbf{dv} = n_{e0}T_{e0}\delta u_{\parallel e}$ and $\int \mu v_{\parallel} B_0 \delta f_e \mathbf{dv} = 3n_{e0}T_{e0}\delta u_{\parallel e}$.

Substituting Eqs. (13) and (27)-(31) into Eq. (A1), we can obtain the Eq. (26) for $\delta A_{\parallel}^{NA}$.